# A Novel VLSI Architecture of Fixed-complexity Sphere Decoder

Bin Wu, Guido Masera
VLSI Lab, Electronics Department
Politecnico di Torino
Turin, Italy
Email: bin.wu@polito.it, guido.masera@polito.it

*Abstract*—Fixed-complexity sphere decoder (FSD) is a recently proposed technique for multiple-input multiple-output (MIMO) detection. It has several outstanding features such as constant throughput and large potential parallelism, which makes it suitable for efficient VLSI implementation. However, to our best knowledge, no VLSI implementation of FSD has been reported in the literature, although some FPGA prototypes of FSD with pipeline architecture have been developed. These solutions achieve very high throughput but at very high cost of hardware resources, making them impractical in real applications. In this paper, we present a novel four-nodes-per-cycle parallel architecture of FSD, with a breadth-first processing that allows for short critical path. The implementation achieves a throughput of 213.3 Mbps at 400 MHz clock frequency, at a cost of 0.18 mm$^2$ Silicon area on 0.13μm CMOS technology. The proposed solution is much more economical compared with the existing FPGA implementations, and very suitable for practical applications because of its balanced performance and hardware-complexity; moreover it has the flexibility to be expanded into an eight-nodes-per-cycle version in order to double the throughput.

## I. INTRODUCTION

Multiple-input multiple-output (MIMO) system has been widely investigated to provide high data-rate and robust wireless link, with an acceptable implementation complexity [1]. It has been already included in some wireless communication standards, such as IEEE 802.16. One of the most challenging problems in MIMO system is to separate the interferences caused by the multiple antennas. Therefore, several MIMO detection algorithms have been proposed to solve this problem.

Among the large variety of MIMO detection techniques, Sphere decoder (SD) is one of the most promising solutions. The well known depth-first sphere decoder employs Schnorr-Euchner enumeration [2] (SEE-SD) to perform tree traversal and achieves maximum-likelihood (ML) performance. However, a major limitation of SEE-SD is the intrinsic variable throughput, which tends to drop off significantly with decreasing signal-to-noise ratio (SNR) [3]. Some other sub-optimal algorithms, such as K-best algorithm and fixed-complexity sphere decoder (FSD), are proposed to obtain constant throughput and lower hardware-complexity, at an acceptable cost of performance loss [4][5]. The first proposed SDs were hard-output, while, recently, soft-output MIMO detectors have been investigated to construct iterative decoding systems integrated with channel decoders, such as convolutional decoders or Turbo decoders, to achieve near-capacity performance [6]. Soft-information could be easily obtained by extending the existing hard-output SD into list sphere decoder (LSD), which generates a candidate list instead of the only ML solution. Then log-likelihood-ratios (LLR) are calculated from the list for each codeword bit and forwarded to the channel decoder.

The soft-output SEE-SD achieves optimal performance, but still with a variable throughput [7]. Efficient VLSI implementations have also been recently proposed for single tree search SD (STS-SD) [8][9], which provide excellent throughput at medium to high SNR values, but tend to be much less efficient at low SNR. Furthermore, several MIMO detection algorithms can hardly be mapped to highly parallel architectures, because of the adopted sequential search order.

Some other algorithms guarantee constant throughput at the price of a certain performance loss, which is due to the use of sub-optimal search methods. Examples of this approach are the soft-output K-best SD [4], which requires sorting operation, and FSD [5], which also achieves a constant throughput with a relatively lower hardware complexity, making it very suitable to perform soft-output MIMO detection. The most attracting advantage of the FSD is the regular tree traversal order that enables highly-efficient implementation. At the same time, the data dependency between two levels of the tree can be avoided in FSD, so allowing for higher clock frequency.

The present work deals with this kind of constant throughput approach and specifically with FSD implementation. As far as we know, all the reported implementations of FSD are based on FPGA devices, and no VLSI solution is available. These implementations achieve very high throughput by employing pipeline architectures, but at the cost of large hardware resources [10][11]. Furthermore, they do not fully exploit the breadth-first visiting order of FSD, in order to improve the decoding speed.

In this paper we present a high-speed and low-cost VLSI implementation of FSD, applying two major innovations:

- A four-nodes-per-cycle parallel architecture that increases the throughput with respect to the usual one-node-per-cycle solution.
- A processing schedule that combines breadth-first node visiting order and pipelining to remove the data dependency between two adjacent levels and shorten the critical path.

These new solutions do not result into an efficient FPGA implementation, mainly because of the poor performance of multi-operand adders on FPGA technology. However it will be shown that the proposed solution enables high efficiency in Silicon implementation. The occupied Silicon area is close to 0.18 mm$^2$, on a 0.13μm CMOS technology, and the achieved throughput is 213.3 Mbps at 400 MHz clock frequency. It is one of the most efficient FSD implementations among the reported works and is easily scalable to higher degrees of parallelism to meet the increasing throughput requirements of current and future wireless communication standards.

This paper is organized as follows: Section II introduces the system model of FSD; details of the proposed four-nodes-per-cycle parallel architecture are discussed in Section III; Section IV compares the overall performance among several implementations of SD; finally, Section V concludes the paper.

## II. SYSTEM MODEL OF FSD

In an iterative MIMO decoding system (as shown in Fig. 1) with $N_t$ transmit antennas and $N_r$ receive antennas, the source bits $x_2$ are firstly encoded into $y_2$ by a channel encoder, such as convolutional encoder or Turbo encoder. Then the coded bit stream after interleaving (Π), $x_1$, is mapped to a $N_t$-dimensional transmit signal vector $s = [s_{Nt-1},...,s_1,s_0]^T$. Each symbol $s_i$ is chosen independently from a complex constellation Ω with $M$ binary bits per symbol, i.e., $|Ω| = 2^M$. For 16-QAM modulation, $M = 4$ and $|Ω| = 16$. The received vector can be denoted as

$$y = Hs + n, \quad (1)$$

where $H$ is the $N_r \times N_t$ complex channel matrix, assumed to be perfectly known at the receiver through channel estimation, $n = [n_{Nr-1},...,n_1,n_0]^T$ is a $N_r$-dimensional complex Gaussian noise vector.

The soft-output MIMO detector is employed to generate soft-information based on the received vector $y$ and the channel matrix $H$. Several algorithms are investigated to perform soft-output MIMO detection, such as minimum mean square error (MMSE) and LSD. In this paper we assume LSD is employed, which performs tree traversal and generates a candidate list in order to calculate LLR. Soft-information is exchanged between the inner MIMO detector and outer channel decoder, through interleaver and deinterleaver [6]. Hard decisions can be made after a number of iterations, depending on the required throughput or bit-error-rate (BER) performance. Then the decoded bit stream is available.

In the transmitted vector symbol constellation $Ω^{N_t}$, there exists a ML solution that can be expressed as

$$s^{ML} = \arg \min_{s \in Ω^{N_t}} \| y - Hs \|^2. \quad (2)$$

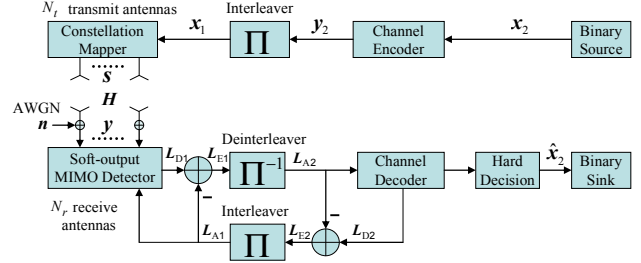

Figure 1. Iterative MIMO decoding system. Subscript '1' denotes variables associated with the inner code and subscript '2' denotes variables associated with the outer channel code.

For a small MIMO system, such as a 4×4 system with QPSK modulation, $s^{ML}$ can be obtained through exhaustive-search [12]. However, it is impractical to perform exhaustive-search for a large system, such as for example a 4×4 system with 16-QAM modulation [13]: in this case a very large number of operations per symbol are required to perform exhaustive search, and this implies the allocation of excessive hardware resources. The SEE-SD is therefore proposed to reduce the complexity, by searching in a hypersphere centered around the received point $y$. The algorithm can be transformed into a tree traversal procedure; proper radius update rules can be adopted to prune the tree, so reducing the computational complexity.

For simplicity, the complex channel matrix can be transformed into real values

$$\hat{H} = \begin{bmatrix} R\{H\} & -I\{H\} \\ I\{H\} & R\{H\} \end{bmatrix}, \hat{y} = \begin{bmatrix} R\{y\} \\ I\{y\} \end{bmatrix}, \quad (3)$$

where $R\{x\}$ and $I\{x\}$ denote the real and imaginary parts of $x$.

Then QR decomposition is applied as

$$\hat{H} = QR, \quad (4)$$

where $Q$ is $2N_r \times 2N_t$ orthogonal matrix, $R$ is $2N_t \times 2N_t$ upper triangular matrix with positive diagonal elements.

Therefore, the partial Euclidean distance (PED) is given by

$$d_i = d_{i+1} + |e_i|^2, i = 2N_t - 1, 2N_t - 2,...,0, \quad (5)$$

where

$$e_i = y_i^{ZF} - \sum_{j=i}^{2N_t-1} R_{i,j}\hat{s}_j = b_i - R_{i,i}\hat{s}_i, \quad (6)$$

$$y^{ZF} = Q^H \hat{y}, \quad (7)$$

$$b_i = y_i^{ZF} - \sum_{j=i+1}^{2N_t-1} R_{i,j}\hat{s}_j, \quad (8)$$

$$\hat{s} = [\hat{s}_{2N_t-1},...,\hat{s}_1,\hat{s}_0]^T = [R\{s\} \quad I\{s\}]^T. \quad (9)$$

At the beginning of decoding, the radius is set to infinity. Tree traversal moves from top to down. Whenever a leaf in the lowest level is reached, and the PED of the leaf is less than the current radius, the radius is replaced by the value of the PED. The SEE-SD performs tree pruning by comparing the PED of each node with the current radius. If the PED is larger than the radius, the node with the whole branch under it is pruned. So the total number of visited nodes is significantly reduced with respect to exhaustive-search, still yielding ML performance.

The soft-output SEE-SD performs the similar depth-first tree search to get a candidate list. The difference compared with the hard-output SEE-SD is that the radius will not be updated until a required number of candidates are obtained [6], making it slower than the hard-output counterpart, because it needs to visit more nodes. In the soft-output case, the whole candidate list is sent to a LLR generator to calculate the soft-information, which is needed by the channel decoder in the iterative MIMO decoding system. LLRs are evaluated for each bit according to

$$L_E(x_k \mid y) \approx \frac{1}{2} \max_{x \in X_{k,+1}} \left\{ -\frac{1}{\sigma^2} \|y - Hs\|^2 + x_{[k]}^T \cdot L_{A,[k]} \right\}$$

$$-\frac{1}{2} \max_{x \in X_{k,-1}} \left\{ -\frac{1}{\sigma^2} \|y - Hs\|^2 + x_{[k]}^T \cdot L_{A,[k]} \right\}, \quad (10)$$

where $X_{k,+1}$ and $X_{k,-1}$ represent the sets of vector $x$ having $x_k = +1$ and $x_k = -1$ respectively, $\sigma^2$ is the noise variance, $x_{[k]}^T$ is the subvector of $x^T$ omitting the bit $x_k$, and $L_{A,[k]}$ is the subvector of the *a-priori* information vector $L_A$ with omitted $L_{A,k}$ [6].

The FSD is also based on PED calculation while traversing the tree. The main difference compared with the SEE-SD is that the FSD does not need to update the hypersphere radius in order to perform tree pruning. At each level of the tree, the number of nodes to be visited is pre-fixed. It can be defined in different ways, which yield different performance. A theoretically derived formula giving the best node distribution along the tree is not available. However, the best solution can be found based on the Monte Carlo simulations for the SEE-SD, and choosing the node numbers in each level from a small set $\{1, N_b\}$, where $N_b$ is the number of branches per node [14]. We performed design space exploration in order to search the best node distribution, and finally found that for the system with 4×4 antennas and 16-QAM modulation, the node distribution of {11111144} achieves the best BER performance. This representation means that in the highest two levels, all the 4 child nodes are visited, while in the lower levels, only one child node with minimum PED is visited. Therefore, 4 nodes are visited in the top level and 16 nodes are visited in each of the lower levels. We also found that a simple sorted QR decomposition (SQRD) [15] significantly improves the performance. Fig. 2 shows the performance offered by both FSD and SEE-SD soft-output detectors coupled with a four state, 1/3 code rate Turbo decoder, which executes 8 decoding iterations; the detectors operate on a 4×4 MIMO channel with 16-QAM modulation and performance are given after two iterations between detection stage and channel decoder. In the reported results, it can be seen that the FSD with SQRD achieves a better performance compared with ordinary QRD.

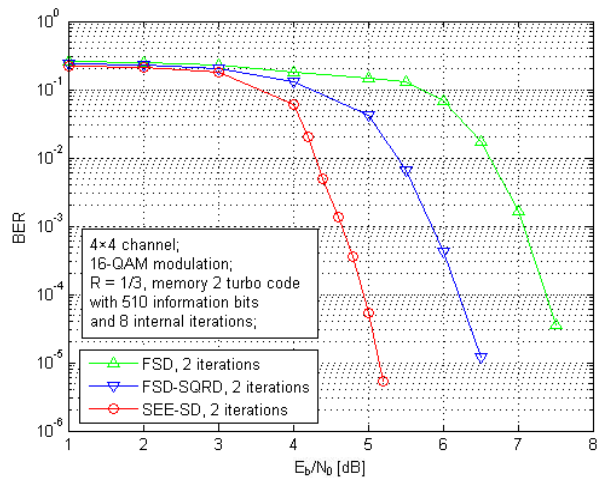

Figure 2. BER performance of the iterative MIMO decoding system for soft-output SEE-SD and FSD with different QR decompositions. The list size is 16 for both SEE-SD and FSD. Node distribution {11111144} is adopted for FSD. "2 iterations" means two iterations between the soft-output MIMO detector and the Turbo decoder.

Because the FSD does not need to reach the lowest tree level immediately to update the radius, the traversal order is very flexible, and can be done either in depth-first or in breadth-first style. We found that the breadth-first order is an essential feature of FSD because the data dependency between a pair of parent-child nodes can be avoided by performing breadth-first tree traversal. It is helpful to shorten the critical path, and we utilized this feature in the proposed four-nodes-per-cycle architecture. Another advantage of the FSD is that the numbers of visited nodes are the same for both hard-output and soft-output versions. Therefore they have the same throughputs.

### III. PARALLEL FSD ARCHITECTURE

Fixed and regular traversal order makes the FSD very suitable to adopt parallel architectures and to improve the throughput. FSD algorithm also admits a scalable amount of parallelism, which enables to trade-off complexity for performance. A straightforward way to increase throughput is to adopt a pipeline architecture, as reported in [10] and [11]. These implementations achieve very high throughput, however, they also involve very high hardware complexity, which makes them impractical in real applications. To solve this problem, we propose a novel four-nodes-per-cycle parallel architecture to reduce the hardware complexity, while maintaining a throughput high enough for most applications.

#### A. Four-nodes-per-cycle architecture

In each level, a group of four nodes are processed in parallel, thanks to the fact that there is no data dependency between them. In the top level, all the four nodes are processed in one cycle, while in each of the lower levels, four cycles are needed to process the 16 nodes which are chosen to be visited, by applying the breadth-first order, in a zig-zag fashion, as shown in Fig. 3. Four nodes of each group in dashed blocks are processed in parallel.

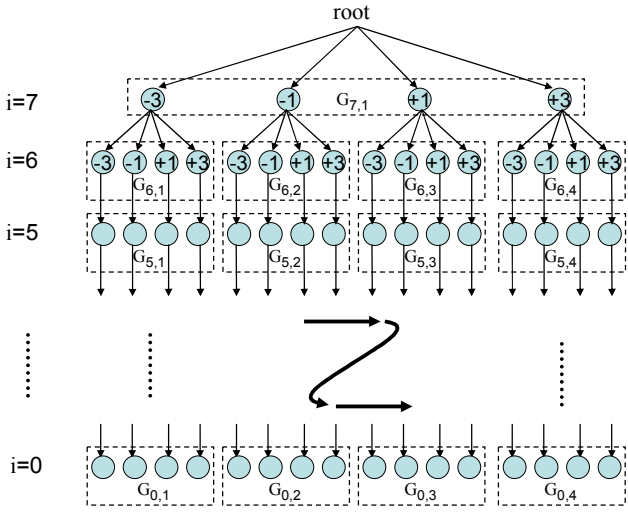

Figure 3. Tree traversal of the four-nodes-per-cycle architecture: each group of four nodes (denoted as $G_{level,column}$) in dashed blocks are processed parallel and in the breadth-first order.

The proposed four-nodes-per-cycle FSD architecture can reach high hardware-efficiency because of the regularity of the tree traversal order. Furthermore, the data dependency between a parent node and its child nodes in DSD can be avoided, making it possible to reach a higher speed.

### B. Breadth-first processing order

Besides the four-nodes-per-cycle method, we also exploited the breadth-first visiting order to shorten the critical path in the FSD architecture. In a SEE-SD, there always exists a data dependency between two of the three main arithmetic tasks: $b_i$ calculation, child node enumeration and PED calculation (see [3]). Fortunately, the data dependency between these arithmetic tasks can be avoided in FSD, by executing at each clock cycle the three tasks on three different groups of four nodes. The critical path is therefore divided into three independent shorter paths. This is an essential advantage of FSD. But it is not exploited in the recently published works such as [10][11]. In these works, nodes in two adjacent levels are processed sequentially and this prevents the three mentioned tasks from being performed in three separated cycles.

In the breadth-first processing order, when a new node is visited, $b_i$ is always ready for use. The PED ($d_i$) of this node can be calculated immediately. In the same cycle, $b_{i-1}$ is calculated synchronously and replaces $b_i$ in memory. Then in the following cycle, direct enumeration is performed for this node to choose a child in the next level. The PED of the chosen node in the next level ($d_{i-1}$) will be calculated after 3 cycles when the same column is processed again. The essence of breadth-first visiting order is that the three main arithmetic tasks, including $b_i$ calculation, $d_i$ calculation and direct enumeration, are processed in the same single cycle for three different groups of nodes, and there is not any data dependency among them, which insures that the critical path is successfully divided into independent shorter paths.

To clarify the proposed breadth-first visiting order, the timing details for the graph example of Fig. 3 are explained in Table I. In each cycle, $d_i$ calculation, $b_i$ calculation and direct enumeration are performed separately for three different groups of node illustrated in Fig. 3. In the first cycle, $b_i$ for all the four groups in level 6 are calculated, because all the nodes in each of the four groups share the same value of $b_i$. Therefore they can be calculated concurrently by the four $b_i$ units. Moreover, in the top two levels, the output of DE units are discarded, because all the nodes in the two levels are visited and it is not needed to perform enumeration.

### C. Architecture details

The implementation is targeted to a MIMO system with 4×4 antennas and 16-QAM modulation. The internal data width is chosen to be 12 bits, based on the evaluation of numerical simulation. In this paper we mainly focus on the FSD implementation. The QR decomposition and LLR calculation are implemented in standard algorithms without special optimization.

#### 1) Diagram of the four-nodes-per-cycle architecture

The block scheme of the four-nodes-per-cycle FSD architecture is shown in Fig. 4. Each of the main arithmetic tasks employs four units with the same internal structure, in order to process four nodes in parallel per clock cycle.

Because of the breadth-first processing order, different tasks of processing a node are performed in different cycles. The signals with index *crt* are referred to the current cycle, while those with index *prv* are related to the previous cycle. The signal lines with slashes are referred to four different values for each of the four nodes in a certain group.

The tree traversal paths in FSD are highly regular, as shown in Fig. 3, which insures that the nodes being processed in each cycle have definite positions. The FSD performs different tasks in each cycle according to the position, which is controlled by a level counter and a column counter in the control unit. The tree traversal path contains eight levels and four columns, as shown in Fig. 3. Therefore a 3-bits register is enough for the level counter and a 2-bits register is employed as the column counter.

#### 2) Calculation of $b_i$

In the first level, $b_7 = y_7^{ZF}$, therefore the value of $y_7^{ZF}$ is directly written into $b_i$ cache at reset. Then in the first cycle, it is immediately used by the $d_i$ units for calculating the PED. In the following levels, the four $b_i$ units are responsible for $b_i$ calculation.

TABLE I.  TASK DISTRIBUTION IN EACH CYCLE

| Cycle | $d_i$ | $b_i$ | DE |
|---|---|---|---|
| 1 | $G_{7,1}$ | $G_{6,1} \sim G_{6,4}$ | - |
| 2 | $G_{6,1}$ | $G_{5,1}$ | - |
| 3 | $G_{6,2}$ | $G_{5,2}$ | $G_{5,1}$ |
| 4 | $G_{6,3}$ | $G_{5,3}$ | $G_{5,2}$ |
| 5 | $G_{6,4}$ | $G_{5,4}$ | $G_{5,3}$ |
| 6 | $G_{5,1}$ | $G_{4,1}$ | $G_{5,4}$ |
| 7 | $G_{5,2}$ | $G_{4,2}$ | $G_{4,1}$ |
| 8 | $G_{5,3}$ | $G_{4,3}$ | $G_{4,2}$ |
| … | … | … | … |

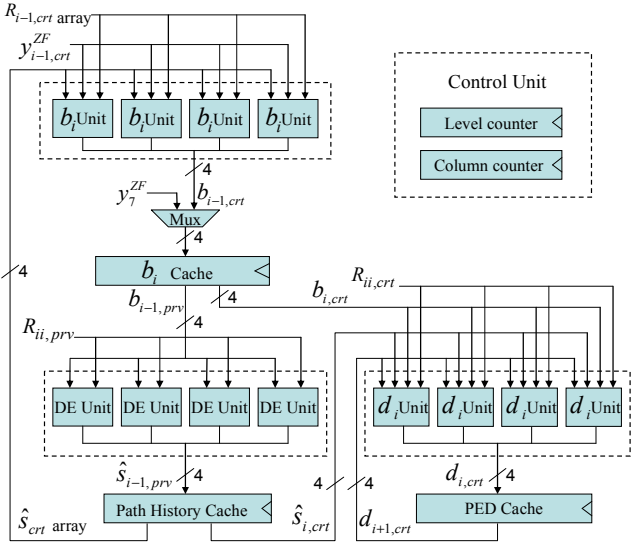

Figure 4. Diagram of the four-nodes-per-cycle FSD architecture

When computing $b_i$, because the value of $\hat{s}_i$ is chosen from a small set $\{+3, +1, -1, -3\}$ for 16-QAM modulation, the multiplication between $R_{i,j}$ and $\hat{s}_i$ can be transformed into an addition:

$$R_{i,j} \times (+3) = R_{i,j} + 2R_{i,j}, \quad (11)$$

$$R_{i,j} \times (+1) = R_{i,j}, \quad (12)$$

$$R_{i,j} \times (-1) = -R_{i,j}, \quad (13)$$

$$R_{i,j} \times (-3) = -R_{i,j} - 2R_{i,j}. \quad (14)$$

The value of $2R_{i,j}$ and $-2R_{i,j}$ can be easily obtained through left shifting operation. But the additions are not performed immediately. Instead, the summands and addends are just left separated. Then a Wallace compress tree [16] is constructed involving all the outputs of the multipliers. In order to shorten the delay path, all of the 14 variables are compressed into a Wallace tree of carry save adder (CSA) with 6 levels, which is followed by a common ripple carry adder (RCA), as shown in Fig. 5.

The $b_i$ units are the most complicated blocks in the FSD architecture. Therefore we considered another solution for $b_i$ calculation in order to choose a more efficient solution. Instead of calculating the values in a single cycle, we distribute the accumulation task into several cycles by employing additional registers to store the updated values of $\sum_{j=i+1}^{7} R_{i,j}\hat{s}_j$ in each cycle, as shown in Fig. 6. Because $R_{6,7}\hat{s}_7$ is used immediately, it does not need to be stored. Therefore 6 registers are needed for lower levels. The output selection is controlled by the level counter in the control unit, as shown in Fig. 4.

We compared the two solutions in order to choose the most efficient one. From synthesis results given in Table II, the complexity of the CSA based $b_i$ calculation is 2,677 GE (gate equivalent). Synthesis has been performed at the clock frequency of 500 MHz on 0.13 μm CMOS technology. As four $b_i$ need to be calculated in parallel, the whole cost is 2,677 × 4 = 10.7 K GE. On the other side, the second solution of Fig. 6 results into a cost of 2,728 GE when synthesized on the same technology at the same clock frequency. The performances of the two individual solutions are quite similar. However if we consider the whole design, the adoption of the second solution in a breadth-first architecture implies that 16 $b_i$ calculation units are allocated, and therefore the total cost is 2,728 × 16 = 43.6 K GE. We conclude that the CSA based solution is more appropriate for breadth-first FSD algorithm.

We also found that the first solution is well suitable for VLSI implementation but not for FPGA, because of its larger number of logic levels, which has significant impact on speed. This will be discussed in Section IV.

*3) Direct enumeration*

Direct enumeration is employed to choose which child node in the next level will be visited, by comparing $|e_i|$ among all the child nodes of a common parent node. The child node with minimum $|e_i|$ is chosen as survival. Because all the nodes are visited in the top two levels, there is no need to perform the direct enumeration. Therefore the output of the DE unit is discarded for nodes belonging to the top two levels.

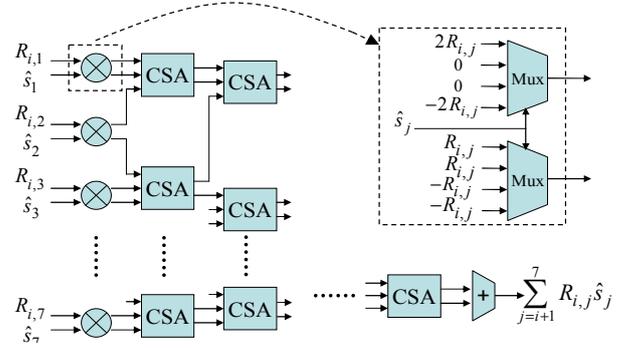

Figure 5. The first solution for $b_i$ calculation

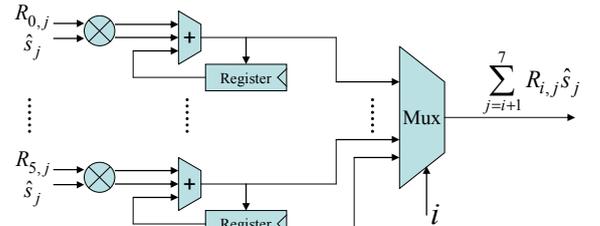

Figure 6. The second solution for $b_i$ calculation

TABLE II. COMPARISON BETWEEN THE TWO SOLUTIONS

| Silicon Area | Solution 1 | Solution 2 |
|---|---|---|
| Combinational area (GE) | 2,677 | 2,000 |
| Noncombinational area (GE) | - | 728 |
| Total area (GE) | 2,677 | 2,728 |

Since $b_i$ is already calculated in the previous cycle, the task of DE unit is just to compare the value of $|b_i - R_{i,i}\hat{s}_i|$ among the child nodes, as shown in Fig. 7. $R_{i,i}$ is multiplied with each element of the set $\{+3, +1, -1, -3\}$ separately. Then all the products are subtracted by $b_i$. Two levels of 12-bits comparators are employed to choose the child node with minimum absolute value of $|b_i - R_{i,i}\hat{s}_i|$.

*4) Calculation of $d_i$*

Four $d_i$ units are employed to calculate the PED of each node. The value of $|e_i|$ has been previously calculated when performing direct enumeration for this node, however it costs more Silicon area to save the value in memory than to calculate it again with the enumerated $\hat{s}_i$. Through VLSI synthesis, we find that the area of a register is similar with the area of a full adder of the same length. However, 16 registers will be needed (the same as the number of candidates in the list) if we want to keep the value of $|e_i|$ for reuse, but if we calculate it again, only 4 adders are necessary.

In Fig. 8, two multipliers are employed, denoted as M1 and M2. M1 is a simple multiplication-addition converter as shown in Fig. 6, while M2 is a common multiplier for calculating the square value of $|e_i|$. The product is then added to the PED of the parent node, i.e., $d_{i+1}$. But only 12 bits of the result are assigned to $d_i$. In the case that the product exceeds the maximum permissible value, $d_i$ is simply set to the maximum permissible value.

*5) Memory organization*

Whenever a new node is chosen to be visited, $\hat{s}_i$ of the node is stored in the path history cache, which will contain the final candidate symbols in the end of decoding. Because all the nodes in the top two levels are visited, their paths have fixed values, which can be directly hard-connected to wires to save Silicon area. Therefore, only the paths in the lowest 6 levels need to be kept in memory. Totally $16 \times 6 \times 2 = 192$ flip-flops are required by the path history cache.

Because of the breadth-first processing order, calculated $b_i$ are not immediately used. They are stored in the $b_i$ cache with 16 entries, the same as the number of candidates in the list.

The PED cache contains the PED of each visited node, which are used to calculate soft-information by LLR generator after finishing the tree traversal. The number of entries is the same as the number of candidates in the list.

The sizes of major registers are given in Table III, in terms of number of flip-flops (number of entries × data width).

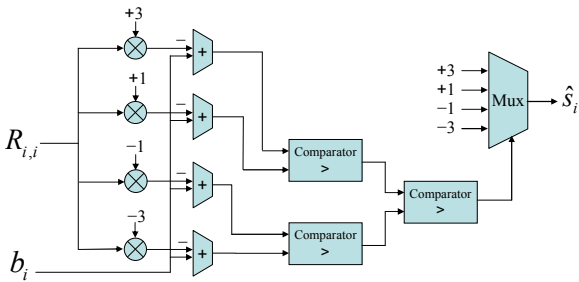

Figure 7. Direct enumeration

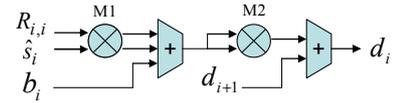

Figure 8. $d_i$ calculation

TABLE III. SIZE OF REGISTERS

| Path history cache | $b_i$ cache | PED cache |
|---|---|---|
| 16×6×2 = 192 | 16×12 = 192 | 16×12 = 192 |

## IV. IMPLEMENTATION RESULTS

### A. Synthesis report

The four-nodes-per-cycle architecture is implemented in VHDL, and is validated with ModelSim. It is then imported into Synopsys Design Vision to perform synthesis, on a 0.13 μm CMOS technology. The synthesis reports show that the implementation of FSD can work at a clock frequency of 400 MHz, with a Silicon area of 0.18 mm². Area breakdown is shown in Table IV. We can see that the $b_i$ units occupy a large portion of the overall area while the DE units and $d_i$ units are relatively smaller.

The four-nodes-per-cycle architecture employs extra three units for each of the major arithmetic tasks, while requires the same number of registers compared with the one-node-per-cycle version. Also the control unit has approximately the same complexity in the two cases. Therefore, we can roughly estimate that the implementation of the one-node-per-cycle version would cost a Silicon area of

$$(42.2\% + 16.7\% + 10.0\%) / 4 + 31.1\% = 48.3\%$$

compared with the four-nodes-per-cycle architecture. It means that the throughput is increased to four times after adopting the four-nodes-per-cycle parallel architecture while the hardware cost is only doubled, showing that the efficiency is improved significantly (93.2% in terms of the throughput/area quotient).

Thanks to the use of the CSA tree, the critical path after synthesis on the 0.13 μm technology dose not lie in the $b_i$ unit. The two levels of 12-bit comparators in the DE unit are the architecture bottleneck in terms of delay and set the maximum clock frequency.

### B. Throughput

The throughput can be denoted as

$$throughput = \frac{f_c \times M \times N_t}{N_c}, \quad (15)$$

where $f_c$ is the clock frequency, $N_c$ is the number of cycles required to perform an entire tree traversal.

TABLE IV. AREA BREAKDOWN

| $b_i$ unit×4 | DE unit×4 | $d_i$ unit×4 | Registers and control | Total area |
|---|---|---|---|---|
| 42.2 % | 16.7 % | 10.0 % | 31.1 % | 100% |

In our implementation for 16-QAM and 4×4 antennas, a total of 4 + 16×7 = 116 nodes are visited in 29 cycles, an additional clock cycle is needed to start a new tree traversal, thus $N_c$ = 30 and the throughput is

$$\frac{400 \times 4 \times 4}{30} = 213.3 \, \text{Mbps}.$$

In the system of 4×4 antennas, with 16-QAM modulation, 16 information bits are transmitted in each channel use. Therefore the throughput without consideration of clock frequency can be given by $(4\times 4)/N_c$ = 0.53 bit/cycle.

### C. Comparison with other works

To evaluate the performance and efficiency of the implemented parallel FSD, we choose two recently published FSD implementations [10][11] and three implementations of other algorithms [3][4][8] for comparison, as shown in Table V. Both the FPGA implementations employ pipeline architectures. Although they are hard-output, the tree traversal parts are similar with the soft-output FSD. The throughputs for FSD and K-best algorithms are all constant, while the throughput for depth-first SD is variable therefore the value is given at 20 dB and 12 dB SNR for [3], at 16 dB and 12 dB SNR for [8].

To facilitate comparisons, the Silicon area is converted into number of Gate Equivalent (GE). The Silicon area is divided by 6.05 ×10$^{-6}$ mm$^2$ (the area of two-input NAND gate with minimum drive strength in the technology library) to obtain GE. However, the hardware-complexity of FPGA implementation is difficult to be compared with Silicon area. So we re-synthesized the four-nodes-per-cycle parallel architecture on Xilinx XC2VP70 FPGA. Only basic hardware resources such as slices, flip-flops and LUTs are used, without particular optimization.

We can see that the speed of the four-nodes-per-cycle architecture on FPGA is much lower than the other two FPGA implementations. The reason for such a large difference is that the critical path after FPGA synthesis is in the $b_i$ unit and it is due to the Wallace tree structure that is very inefficient for FPGA implementation. It has 24 logic levels, leading to a long delay path. However, the FPGA results are only used to compare the hardware-complexity.

The two FPGA implementations both employ pipeline architectures. They achieve very high throughput at relatively lower clock frequency. [10] visits all nodes in one level in each clock cycle, with a throughput of 4 bit/cycle, and [11] uses two cycles to complete the same task, with a throughput of 2 bit/cycle. They are much higher than the proposed four-nodes-per-cycle architecture, which achieves only 0.53 bit/cycle. However, the high throughput comes with very high hardware complexity, making them impractical for real applications.

The hardware cost of the proposed four-nodes-per-cycle architecture on FPGA is much smaller than the other two FSD implementations employing pipeline architectures, even if no particular optimizations for FPGA are applied and no special hardware resources such as multipliers and DSP blocks are utilized. The VLSI implementation is also compact and costs only 29.8 K GE on Silicon.

The soft-output K-best implementation MKSE in [4] achieves an acceptable throughput, however, at a high cost of hardware resource. Furthermore, the clock frequency reaches to only 200 MHz, on a 0.13μm CMOS technology. The hard-output SEE-SD implementation ASIC-II in [3] achieves higher throughput at 20 dB SNR and costs lower hardware resource compared with MKSE. However, the throughput is variable, and drops dramatically at lower SNR. For example, the throughput drops to 85.9 Mbps at 12 dB SNR and to less than 50 Mbps when the SNR is below 5 dB. Therefore the FSD and the K-best algorithms are more efficient compared with the depth-first SEE-SD in low SNR conditions, because of the constant throughput. The soft-output STS-SD implementation reported in [8] also achieves a variable throughput at a higher cost of hardware-complexity than the four-nodes-per-cycle architecture.

Although the other two FSD implementations achieve higher throughput, they also cost much higher hardware resources, which is definitely a burden for practical applications. Instead of pursuing high throughput, the four-nodes-per-cycle parallel architecture aims at reducing the hardware-complexity while maintaining a relatively high throughput for most applications. Furthermore, the proposed four-nodes-per-cycle architecture is very flexible to be extended into an eight-nodes-per-cycle version in order to double the throughput.

TABLE V.    COMPARISON BETWEEN SD IMPLEMENTATIONS

| Work | Four-nodes-per-cycle FSD | | [10] | [11] | [4] | [3] | [8] |
|---|---|---|---|---|---|---|---|
| Antennas | 4×4 | | 4×4 | 4×4 | 4×4 | 4×4 | 4×4 |
| Modulation | 16-QAM | | 16-QAM | 16-QAM | 16-QAM | 16-QAM | 16-QAM |
| Algorithm | FSD | | FSD | FSD | K-best | Depth-first SD | STS-SD |
| List Size | 16 | | 16 | 16 | 5 | - | - |
| Technology | 0.13 μm CMOS | FPGA XC2VP70 | FPGA XC2VP70 | FPGA EP2S60F672C3 | 0.13 μm CMOS | 0.25 μm CMOS | 90 nm CMOS |
| Hardware Cost | 29.8 K GE | 3,458 slices (10%) 660 flip-flops (1%) 6,587 LUTs (9%) | 12,721 slices (38%) 15,332 flip-flops (23%) 16,119 LUTs (24%) 160 multipliers (48%) 82 RAM blocks (25%) | 13,743 ALUTs (28.2%) 1,412 flip-flops (2.94%) 4 DSP blocks | 97 K GE | 50 K GE | 60 K GE |
| Max. Clock Freq. | 400 MHz | 52 MHz | 100/150 MHz | 102 MHz | 200 MHz | 71 MHz | 384 MHz |
| Throughput | 213.3 Mbps | 27.7 Mbps | 400/600 Mbps | 800 Mbps | 106.6 Mbps | 169 Mbps@20 dB 85.9 Mbps@12 dB | 70 Mbps@16 dB 10 Mbps@12 dB |

*D. Real vs. complex channel model*

We adopted real valued channel model to implement the four-nodes-per-cycle FSD architecture. The effects of using real or complex models have been widely discussed in the literature [17]. For SEE-SD, smaller number of tree levels helps speeding up the updating of radius. Therefore the complex model needs to visit lower number of nodes, yielding higher throughput. For breadth-first SD algorithms, such as K-best and FSD, although the number of visited nodes is constant, the real model needs to visit approximately double of the nodes compared with the complex model with the same list size, because of the doubled number of tree levels. However, the higher throughput of complex model comes at the higher cost of hardware-complexity, because the real and the imaginary parts need to be processed concurrently. Therefore appropriate trade-offs are needed for different applications. The study in [17] suggests that the real valued model results into more efficient hardware implementation.

For FSD, complex model leads to higher degree of parallelism, but also results in higher hardware-complexity, as shown in [10][11]. The increased hardware-complexity mainly comes from two factors: approximately doubled arithmetic tasks for real and imaginary parts, and the child node enumeration from a larger number of child nodes. For example, in a system with 16-QAM modulation, the enumeration from four child nodes is relatively simple with a real model. But for complex model it becomes more complicated because the number of child nodes increases to 16. More multipliers and comparators are needed to perform the enumeration, not only increasing the complexity, but also slowing the speed.

To summarize, we can say that the adoption of real or complex channel model results into different trade-offs between throughput and occupied area. Moreover, these trade-offs also depend on the specific chosen algorithm to visit the tree. The choice of the real model in this work is motivated by the search for a low-area implementation.

## V. CONCLUSION

In this paper we present a low cost implementation of the Fixed-complexity Sphere Decoder. FSD is very suitable for constructing iterative MIMO decoding systems and allows for highly efficient parallel architectures. We propose a novel four-nodes-per-cycle parallel architecture that exploits the breadth-first visiting order to improve throughput and shorten the critical path. Compared with the previously published pipeline architectures, the four-nodes-per-cycle architecture reduces the cost of hardware resources significantly while maintains a high throughput, making it suitable for practical applications with balanced performance and hardware-complexity. It is also very flexible to be extended to meet the increasing requirement of new wireless communication standards.

ACKNOWLEDGMENT

The authors would like to thank Maurizio Martina for providing the C++ implementation of the Turbo decoder.

REFERENCES

[1] A. J. Paulraj, D. A. Gore, R. U. Nabar, H. Bölcskei, "An overview of MIMO communications - a key to gigabit wireless," Proceedings of the IEEE, vol. 92, no. 2, pp. 198−218, February 2004.

[2] E. Agrell, T. Eriksson, A. Vardy, K. Zeger, Zeger, "Closest point search in lattices," IEEE Trans. Inf. Theory, vol. 48, no. 8, pp. 2201–2214, August 2002.

[3] A. Burg, M. Borgmann, M. Wenk, M. Zellweger, W. Fichtner, H. Bölcskei, "VLSI implementation of MIMO detection using the sphere decoding algorithm," IEEE J. Solid-State Circuits, vol. 40, no. 7, pp. 1566−1577, July 2005.

[4] Z. Guo, P. Nilsson, "Algorithm and implementation of the K-best sphere decoding for MIMO Detection," IEEE J. Sel. Areas Commun., vol. 24, no. 3, pp. 491–503, March 2006.

[5] L. G. Barbero, J. S. Thompson, "Extending a fixed-complexity sphere decoder to obtain likelihood information for Turbo-MIMO systems," IEEE Trans. Vehicular Tech., vol. 57, no. 5, pp. 2804−2814, September 2008.

[6] B. M. Hochwald, S. ten Brink, "Achieving near-capacity on a multipleantenna channel," IEEE Trans. Commun., vol. 51, no. 3, pp. 389–399, March 2003.

[7] J. Lee and S.-C. Park, "Novel techniques of a list sphere decoder for high throughput," Proceedings of the ICACT, Phoenix Park, Korea, February 2006, vol. 3, pp. 1785–1787.

[8] E. M. Witte, F. Borlenghi, G. Ascheid, R. Leupers, H. Meyr, "A scalable VLSI architecture for soft-input soft-output depth-first sphere decoding," arXiv: 0910.3427, October, 2009.

[9] C. Studer, A. Burg, H. Bölcskei, "Soft-output sphere decoding: algorithms and VLSI implementation," IEEE J. Sel. Areas Commun., vol. 26, no. 2, pp. 290−300, February, 2008.

[10] L. G. Barbero, J. S. Thompson, "Rapid prototyping of a fixed-throughput sphere decoder for MIMO systems," Proceedings of the IEEE ICC, Istanbul, Turkey, June 2006, vol. 7, pp. 3082–3087.

[11] M. S. Khairy, M. M. Abdallah, S. E.-D. Habib, "Efficient FPGA implementation of MIMO detector for mobile WiMAX system," Proceedings of the IEEE ICC, Dresden, Germany, June, 2009, pp. 1–5.

[12] A. Burg, N. Felber, W. Fichtner, "A 50 Mbps 4×4 maximum likelihood decoder for multiple-input multiple-output systems with QPSK modulation," Proceedings of the IEEE ICECS, Sharjah, United Arab Emirates, December, 2003, vol. 1, pp. 332–335.

[13] D. Garrett, L. Davis, S. ten Brink, B. Hochwald, G. Knagge, "Silicon complexity for maximum likelihood MIMO detection using spherical decoding," IEEE J. Solid-State Circuits, vol. 39, no. 9, pp. 1544–1552, September 2004.

[14] J. Jaldén, L. G. Barbero, B. Ottersten, J. S. Thompson, "The error probability of the fixed-complexity sphere decoder," IEEE Trans. Signal Process., vol. 57, no. 7, July 2009.

[15] D. Wübben, R. Böhnke, J. Rinas, V. Kühn, K. D. Kammeyer, "Efficient algorithm for decoding layered space-time codes," Electronics Letters, vol. 37, no. 22, October 2001.

[16] P. Ienne, A. K. Verma, "Arithmetic transformations to maximise the use of compressor trees," Proceedings of the second IEEE international Workshop on Electronic Design, Test and Applications, 2004, DELTA 2004, pp. 219−224.

[17] M. Myllylä, M. Juntti, "Implementation aspects of list sphere detector algorithms," Proceedings of the IEEE Global Telecommunications Conference, 2007, GLOBECOM '07, pp. 3915−3920.